%% file: RD_ISIT_v2.tex
\documentclass[conference]{IEEEtran}
\usepackage{amsmath,amssymb}
\usepackage[mathscr]{eucal}
\usepackage{url}
\usepackage{graphics,graphicx,multicol}
\usepackage{color}
\newcommand{\beq}{\begin{eqnarray}}
\newcommand{\eeq}{\end{eqnarray}}
\newcommand{\n}{\nonumber}
\newcommand{\br}{{\mathbf r}}
\newcommand{\bs}{{\mathbf s}}
\newcommand{\bc}{{\mathbf c}}
\newcommand{\cC}{{\mathcal C}}

\newcommand{\bD}{{\mathbf D}}
\newcommand{\bH}{{\mathbf H}}
\newcommand{\balpha}{\boldsymbol\alpha}
\newcommand{\bbeta}{\boldsymbol\beta}
\newcommand{\blam}{\boldsymbol\lambda}
\newcommand{\bmu}{\boldsymbol\mu}

\title{Network Control: A Rate-Distortion Perspective}

\author{Jubin Jose and Sriram Vishwanath\\
Dept. of Electrical and Computer Engineering\\
The University of Texas at Austin\\
\{jubin, sriram\}@austin.utexas.edu
}

\begin{document}

\maketitle

\begin{abstract}

Today's networks are controlled assuming pre-compressed and packetized data. For video, this assumption of data packets abstracts out one of the key aspects - the lossy compression problem. Therefore, first, this paper develops a framework for network control that incorporates both source-rate and source-distortion. Next, it decomposes the network control problem into an application-layer compression control, a transport-layer congestion control and a network-layer scheduling. It is shown that this decomposition is optimal for concave utility functions. Finally, this paper derives further insights from the developed rate-distortion framework by focusing on specific problems.
\end{abstract}

\section{Introduction}

Traffic forecasts predict mobile video to be a very significant slice ($66$\%) of the world's mobile data by $2014$ \cite{CISCO_VNI_2014}. Video is different from data and voice as video-streaming quality can be varied, which impacts the user experience. Existing network architecture and protocols are primarily designed for data. Hence, majority of the literature assumes an existing packetized system, and then optimizes network performance \cite{shakkottai2007network}. An important component that is absent from such a framework is (lossy) compression. Compression is typically understood as an application-layer operation and thus separated from the network protocol stack optimization. However, the extent and nature of the compression employed critically impacts user experience, especially for video-streaming. Assuming the sources are already quantized/compressed leads to a formulation that presents only a partial picture on the quality of service observed by the users in the system. For instance, lightly-compressed video may require rates much higher than those that can be allocated while ensuring stable network operation, while heavily compressed video, although easy to deliver, reduces the quality of the end-user's experience. Thus, the distortion experienced by each user must be optimized to provide the best user experience (See \cite{chou2006rate,kalman2003rate,chakareski2006rate} and references therein). 

In this paper, we build a framework for network control by applying rate-distortion theory \cite{berger1971rate}. Traditional network control can be viewed as a special case where the distortion (and thus compression algorithm) is fixed at a value independent of network state and overall user utility function. Distributed compression problems have been studied and partially solved for special cases (such as Gaussian and/or binary sources) for particular settings. Typically, there is {\em no} provably optimal separation between source and channel coding in networks. However, for the special case of independent sources being transmitted through the network, it is known that separate source and channel coding is optimal \cite{Tian2010}. In this paper, we focus on independent (uncompressed) sources in the network that must be compressed and subsequently transmitted through the network. This applies to many scenarios including video-streaming. Furthermore, focusing on source-channel separation allows us to develop a rate-distortion framework that scales with the network size, and hence, applicable to large real-world networks.

\subsection{Main results}

For networks with mutually independent (but possibly temporally correlated) sources, we consider the two quantities - (\emph{i}) the \emph{source-entropy} and (\emph{ii}) its \emph{distortion-offset} that are sufficient in representing compression. We formally define these two quantities in Section \ref{sec:RDframework}. Using these, we develop a framework for network control. Few of the important implications of our framework are:
\begin{itemize}
\item With lossy compression, the traditional notion of flow conservation does not hold. This has far-reaching consequences in network protocol design.
\item Our formulation based on \emph{source-entropy} and \emph{distortion-offset} has only linear constraints in addition to capacity constraints. Hence, if we focus on concave utility functions, existing convex optimization techniques can be applied, especially variants of distributed algorithms developed in \cite{kelly2001mathematical}. 
\end{itemize}

Based on the rate-distortion framework, we present the following results:
\begin{itemize}
\item We show optimal decomposition of network control into three layers:  (\emph{a}) an application-layer with compression control, and (\emph{b}) a transport-layer with congestion control, and (\emph{c}) a network-layer with MaxWeight scheduling.
\item For a compression problem with binary sources and proportional-fair like utility functions, we derive the optimal control policy. The optimal policy requires varying distortion based on link-rate, and hence, clearly shows the sub-optimality of decoupling compression.
\item We solve a specific problem involving sending binary uncompressed sources over a Gaussian multiple access channel.
\end{itemize}

\subsection{Organization}
The next section presents the rate-distortion framework for network control. Section \ref{sec:layering} derives an optimal decomposition into layers. Section \ref{sec:compression} studies a compression control problem for binary sources. Section \ref{sec:MAC} applies the framework to Gaussian MACs. The paper concludes with Section \ref{sec:concl}.

\section{A Rate-distortion framework for network control}
\label{sec:RDframework}

We consider a single-hop\footnote{The framework naturally extends to multi-hop networks.} network with $N$ independent sources, labeled $i=1,2,\ldots, N$. The $i$-th (possibly continuous-valued) source $X_i$ has an \emph{uncompressed-rate} of $s_i$ symbols/sec. This source is compressed at a \emph{distortion} of $D_i$ (per symbol, averaged across time) to a \emph{rate} of $c_i$ bits/sec. In other words, a lossy-compression code exists that maps vectors comprised of source symbols to binary vectors such that recovery is  possible to within a distortion of $D_i$ per symbol. Mathematically, a rate-distortion code (operating over blocks of symbols of size $n$, with $n$ large enough) of rate $c_i + \epsilon$ bits/sec exists for source $X_i$ such that reconstruction to within a distortion $D_i$ is possible such that $\epsilon \rightarrow 0$ as $n \rightarrow \infty$.

This compressed source is transmitted over a link with \emph{link-rate} of $r_i$ bits/sec. The corresponding vectors are denoted by $\bs$, $\bD$, $\bc$ and $\br$, respectively. These link rates are coupled in a wireless network, and this, for a single-hop network, is captured by the $N$-dimensional information-theoretic rate region denoted by $\cC$ (This rate region may be the capacity region if the network's capacity region is known, or the best known rate region if unknown). The parameters introduced so far are associated with different functionalities in a network: (\emph{a}) $s_i$ and $D_i$ are associated with (lossy) source-coding, (\emph{b}) $c_i$ is associated with congestion (or rate) control, and (\emph{c}) $r_i$ is associated with rate allocation (or scheduling). 

The source-coding, rate control and scheduling problems are tied to each other closely. As a result, the parameters associated with these problems must be jointly optimized. Therefore, we desire a framework that captures all these problems. However, the traditional framework does not include the source-coding component. It is based on the following optimization problem:
\beq
\label{eq:trad_NUM}
&\max \limits_{\br } &\sum_{i=1}^N U_i(c_i) 
\eeq
subject to
\beq
\label{eq:trad_const1}
&& c_i \le r_i, \forall i, \\
\label{eq:trad_const2}
&& \br \in \cC.
\eeq
In this framework, $U_i(c_i)$ in (\ref{eq:trad_NUM}) is the (concave) utility function associated with the (compressed) rate $c_i$ of $i$-th source and (\ref{eq:trad_const1})-(\ref{eq:trad_const2}) are capacity constraints. This framework can be decomposed into two layers: a transport-layer performing rate control, and a network-layer performing scheduling \cite{palomar2006tutorial}. 

To incorporate the source-coding parameters, it is natural to utilize rate-distortion functions of sources studied in information theory \cite{Cover2006elements}. For explaining this, we consider two source types: 
\begin{enumerate}
\item {\bf Binary sources with Hamming distortion:}  Consider independent Bernoulli($p_i$) binary sources that are mutually independent arriving at rates of $s_i$ symbols/second. The rate-distortion function for this source is known to be
\beq
\label{eq:binary_RD}
R(s_i,D_i) = s_i \left(H(p_i) - H(D_i)\right),
\eeq
where $H(\cdot)$ is the binary entropy function. Now, motivated from (\ref{eq:binary_RD}), we define two variables to represent this source: (\emph{a}) \emph{source-entropy} 
\beq
\label{eq:bin_alph}
\alpha_i = s_iH(p_i)
\eeq in bits/sec, where $s_i$ is the uncompressed-rate in symbols/sec and $0<p_i<1$ is the given Bernoulli parameter of $i$-th source, and (\emph{b}) (negative) \emph{distortion-offset} 
\beq
\label{eq:bin_bet}
\beta_i = -s_iH(D_i)
\eeq in bits/sec, where $D_i$ is the Hamming distortion per symbol. 
\item {\bf Gaussian sources with squared-error distortion:} Consider zero-mean independent Gaussian sources with variances $\sigma_i^2$ arriving at a rate of $s_i$ symbols per second. With squared-error distortion, the rate-distortion function is known to be
\beq
\label{eq:Gaussian_RD}
R(s_i,D_i)=\frac{s_i}{2}\log_2 \frac{\sigma^2_i}{D_i}.
\eeq
Now, differential \emph{source-entropy} $\alpha_i$ and \emph{distortion-offset} $\beta_i$ are defined as follows: 
\[
\alpha_i = \frac{s_i}{2}\log_2 2 \pi e\sigma^2_i,
\]where $\sigma^2_i>0$ is the given variance parameter of the $i$-th source, and
\[
\beta_i = -\frac{s_i}{2}\log_2 2\pi e D_i,
\]where $D_i$ is the squared-error distortion per symbol. Note that these two variables can take both positive and negative values.
\end{enumerate}

Now, \emph{source-entropy} and \emph{distortion-offset} can be identified as two parts of the rate-distortion function for multiple types of sources, both i.i.d. and correlated (for example, see Shannon's rate-distortion lower bound \cite{Cover2006elements}). Denoting source-entropy and distortion-offset as $\alpha_i$ and $\beta_i$ respectively, we have a  tradeoff between the two of the form given by:
\beq
\label{eq:rate_ineq}
\alpha_i + \beta_i \le c_i,\forall i.
\eeq
This simply states that the compressed rate should be higher than the fundamental limit given by the rate-distortion function.  Since distortion-offset terms appear in the constraints, it shows that flow conservation assumed in data networks does not hold for sources such as video. This motivates re-design of network protocol components that assume packets to be immutable. 

Now, a user's happiness (or user experience) can be thought of as a function of the source-entropy and distortion-offset. Therefore, a natural framework for network control is to maximize the sum of the user experience subject to all network constraints. For deriving a suitable layering architecture in the next section, we consider a slightly different looking but equivalent\footnote{$U_i(c_i)$ can be absorbed into $V_i(\alpha_i,\beta_i)$} framework for network control:
\beq
\label{eq:convex_NUM}
&\max\limits_{\balpha,\bbeta,\bc,\br} &\sum_{i=1}^N V_i(\alpha_i,\beta_i) + U_i(c_i) 
\eeq
subject to
\beq
\label{eq:const1}
&& \alpha_i + \beta_i \le c_i, \forall i,\\
\label{eq:linconst1}
&&a_i\alpha_i \ge 0, b_i \beta_i \le 0, \forall i, \\
\label{eq:linconst3}
&&\alpha_i + \beta_i \ge 0, \forall i, \\
\label{eq:linconst4}
&&c_i \le r_i, \forall i, \\
\label{eq:const2}
&& \br \in \cC, 
\eeq
where $a_i,b_i \in \{0,1\}$ are constants that are source-dependent, (\ref{eq:const1})-(\ref{eq:linconst3}) are rate-distortion conditions and (\ref{eq:linconst4})-(\ref{eq:const2}) are capacity constraints. The rate-distortion framework in (\ref{eq:convex_NUM}) has two main advantages. (\emph{i}) It presents a notion of  optimal network optimization while dealing with uncompressed sources.
(\emph{ii}) The constraints in (\ref{eq:const1})-(\ref{eq:linconst4}) are linear, and $\cC$ in (\ref{eq:const2}) is a convex set (with time sharing). Hence, with concave utility functions, we have a convex framework.

\section{Decomposition into Multiple Layers}
\label{sec:layering}

In this section, we show that the framework in (\ref{eq:convex_NUM}) can be decomposed into three layers: (\emph{a}) ``application'' layer with compression control,  (\emph{b}) ``transport'' layer with congestion control, and  (\emph{c}) ``network'' layer with (centralized) scheduling. As evident from the names, each of these layers has direct correspondence with a layer in the standard network protocol stack. 

We proceed by introducing two sets of dual variables. We introduce non-negative dual variables $\mu_i,\forall i$ (vector denoted by $\bmu$) corresponding to constraints in (\ref{eq:const1}), and non-negative dual variables $\lambda_i,\forall i$ (vector denoted by $\blam$) corresponding to constraints in (\ref{eq:linconst4}). With these dual variables, we obtain the following Lagrangian:
\beq
{\mathcal L} &=& \sum_{i=1}^N V_i(\alpha_i,\beta_i) + U_i(c_i) \n\\
\label{eq:lagrangian}
&&- \sum_{i=1}^N \mu_i( \alpha_i + \beta_i-c_i)- \sum_{i=1}^N \lambda_i(c_i-r_i).
\eeq 
Now, the dual objective $g(\bmu,\blam)$ is defined as
\beq
g(\bmu,\blam)=&\max\limits_{\balpha,\bbeta, \bc, \br} &\sum_{i=1}^N V_i(\alpha_i,\beta_i) - \mu_i( \alpha_i + \beta_i) \n \\
&&+ \sum_{i=1}^NU_i(c_i) - (\lambda_i - \mu_i)c_i\n \\ 
\label{eq:NUM_reform}
&&+\sum_{i=1}^N \lambda_ir_i
\eeq
subject to (\ref{eq:linconst1})-(\ref{eq:linconst3}) and (\ref{eq:const2}). From Langrange duality, it is well-known that $g(\bmu,\blam)$ gives an upper bound on the primal problem in (\ref{eq:convex_NUM}) for feasible primal and dual variables. This leads to the dual problem to obtain an upper bound on the primal problem, given by
\beq
\label{eq:NUM_dual}
&\min\limits_{\bmu,\blam} &g(\bmu,\blam)\\
&\text{s.t.}&\lambda_i \ge 0, \mu_i \ge 0, \forall i. \n
\eeq
For concave utility functions, under mild conditions \cite{BV04}, it follows that this dual problem is tight, i.e., the optimal value of (\ref{eq:NUM_dual}) is equal to the optimal value of (\ref{eq:convex_NUM}).

Now, it is fairly straightforward to see that the Lagrangian formulation in (\ref{eq:NUM_reform}) decomposes into the following optimization problems: 
\begin{enumerate}
\item {\bf Distributed Compression Control:} For all $i$, given $\mu_i$,
\beq
\label{eq:RD_Control}
&\max \limits_{\alpha_i,\beta_i} &V_i(\alpha_i,\beta_i)  - \mu_i(\alpha_i + \beta_i) \\
&\text{s.t.}& a_i\alpha_i \ge 0, b_i \beta_i \le 0, \alpha_i + \beta_i \ge 0. \n 
\eeq
\item {\bf Distributed Congestion Control:} For all $i$, given $\mu_i$ and $\lambda_i$,
\beq
\label{eq:rate_control}
&\max \limits_{c_i} & U_i(c_i)-(\lambda_i-\mu_i)c_i.
\eeq
\item {\bf Centralized MaxWeight Scheduling:} Given all $\lambda_i$,
\beq
\label{eq:MaxWeight}
&\max \limits_{\br} &\sum_{i=1}^N \lambda_ir_i\\
&\text{s.t.} & \br \in \cC. \n
\eeq
\end{enumerate}

In contrast to existing network control, the distributed compression problem in (\ref{eq:RD_Control}) is explicitly included in our decomposition. This problem jointly chooses source-entropy and distortion-offset based on the utility function. The congestion control in (\ref{eq:rate_control}) and the centralized scheduling in (\ref{eq:MaxWeight}) match with those known in existing literature \cite{shakkottai2007network, B:GNT06}. Note that all three problems in (\ref{eq:RD_Control}), (\ref{eq:rate_control}) and (\ref{eq:MaxWeight}) are coupled through dual variables $\bmu,\blam.$ In many cases, it is possible to use gradient methods to solve for the dual variables \cite{palomar2006tutorial}. We do not delve into a discussion of such methods to solve these problems. Instead, we focus on two problems to obtain further insights in combining compression with network control - first, we study a compression control problem for binary sources, and then, we apply our framework to Gaussian multiple access channels (MACs).

\section{Compression Control for Binary Sources}
\label{sec:compression}

Let us consider the lossy compression problem that determines source-entropy and distortion-offset given a compressed-rate. We study this problem to understand the tradeoff involved in choosing higher source-rate (with higher distortion) versus lower source-rate (with lower distortion). With utility functions that are strictly increasing, it follows that optimal parameters satisfy $\alpha_i + \beta_i \le c_i$ with equality. Under this setting, the compression control at every source is: for given $a_i$, $b_i$ and $c_i$
\beq
\label{eq:RD_Control_simp}
&\max \limits_{\alpha_i} &V(\alpha_i,c_i-\alpha_i)
\eeq
subject to
\beq
&&a_i\alpha_i \ge 0,\n \\
&& b_i(c_i-\alpha_i) \le 0. \n 
\eeq

In order to obtain explicit solutions to the optimization problem in  (\ref{eq:RD_Control_simp}) , we solve it in the context of binary source with Hamming distortion. For a binary source, we have $a=1$ and $b=1$. Consider the utility function\footnote{This is an example, and the choice of utility functions that is appropriate in practice is a subject for further study.}:
\beq
\label{eq:prop_utility}
V(\alpha_i,\beta_i) = \log_e \alpha_i + K_i \beta_i,
\eeq
for some constant $K_i > 0$. Note that this utility function is an extension of the proportional-fair utility function with linear penalty for distortion-offset. Therefore, (\ref{eq:RD_Control_simp}) simplifies to
\beq
\label{eq:RD_Control_bin}
&\max \limits_{\alpha_i} &\log_e \alpha_i + K_i(c_i-\alpha_i)\\
&\text{s.t.} &\alpha_i \ge c_i.\n 
\eeq
The unconstrained problem in (\ref{eq:RD_Control_bin}) is maximized at $\alpha_i = 1/K_i.$ Therefore, for the constrained problem in (\ref{eq:RD_Control_bin}), we have
\beq
\label{eq:alpha_opt}
\alpha_i^* = \left\{
\begin{array}{l}
1/K_i, \text{ if } 1/K_i \ge c_i\\
c_i, \text{ otherwise.} 
\end{array}
\right.
\eeq
This simple rate-distortion-control policy can be implemented as long as the application layer is aware of the compressed-rate $c_i$. 

The expression in (\ref{eq:alpha_opt}) provides a simple rule to decide whether to transmit at zero-distortion, i.e., with source-entropy $\alpha_i = c_i$ and distortion-offset $\beta_i = 0$, or transmit with distortion, i.e., source-entropy $\alpha_i = 1/K_i$ and distortion-offset $\beta = c_i - 1/K_i.$ When $1/K_i \ge c_i$, substituting $\alpha_i= 1/K_i$ and $\beta_i = c_i - 1/K_i$ in (\ref{eq:bin_alph}) and (\ref{eq:bin_bet}), respectively, we get the following: uncompressed-rate $s$ in symbols/sec is given by
\[
s_i=\frac{1}{K_iH(p)},\]
and Hamming distortion $D_i$ is given by the expression
\[
\frac{H(D_i)}{H(p)} = 1- c_iK_i.
\]
Recall that $p$ is the Bernoulli parameter associated with source and $H(\cdot)$ is the binary entropy function. 

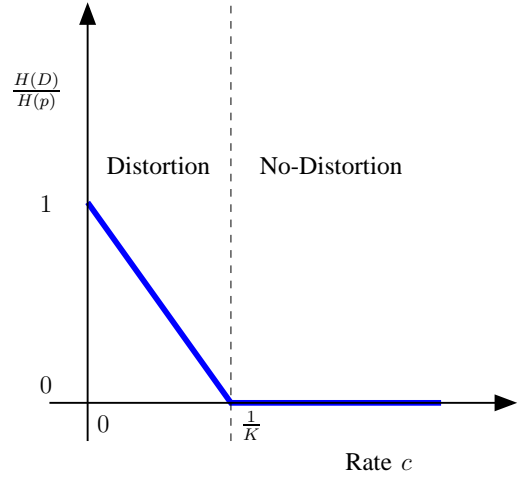
\begin{figure}[!h]
\centering
\scalebox{0.8}{\input{./rd_control.pstex_t}}
\caption{Distributed rate-distortion-control for binary sources; Region to the left of dashed line represents source-coding with distortion and to the right represents source-coding without distortion}
\label{fig:rd_control}
\end{figure}

This compression rule is depicted in Figure \ref{fig:rd_control}. In simple words, this rule states that source-coding with distortion has to be performed at low compressed-rates and source-coding without distortion has to be performed at high compressed-rates. Furthermore, the amount of distortion introduced by the compression algorithm is piecewise linear. This shows that the traditional approach of decoupling compression control from network optimization is sub-optimal. In majority of existing video-streaming systems, compression control is performed using ad hoc algorithms.

\section{Rate-Distortion Framework applied to Multiple Access Channels}
\label{sec:MAC}

Our goal is to understand the interplay between compression and communication - specifically, the way channel capacity and resulting distortion impact  one another. We choose Gaussian multiple access channel (MAC) for our analysis here as they represent the simplest multi-terminal system model, and the capacity region for a MAC is well known \cite{Cover2006elements}. Further, we consider simple utility functions below that are only dependent on the distortion suffered in the compression process. These simplifications help us focus on our goal.

Consider two i.i.d. Bernoulli($p_i$) binary sources that are mutually independent (across sources) arriving at rates of $s_i$ symbols per second. For a binary source with Hamming distortion, the rate-distortion function is given by (\ref{eq:binary_RD}). The uncompressed-rates $s_i$ are positive constants that are fixed by nature and assumed to be known.  After compression, these two sources are to be communicated over a Gaussian multiple access channel. 

Now, network control for this example can be expressed as
\beq
\label{eq:d_control}
&\max \limits_{\bD} &\sum_{i=1}^2 V_i(D_i)
\eeq
subject to
\beq
&& s_i \left(H(p_i) - H(D_i)\right) \le C(P_i), \forall i, \n \\
&& \sum_{i=1}^2 s_i \left(H(p_i) - H(D_i)\right) \le C(P_1+P_2), \n \\
&&D_i \ge 0, D_i \le 1, \forall i. \n 
\eeq
Here, we have used the capacity region of the Gaussian MAC channel. $C(\cdot)$ corresponds to Shannon's capacity formula given by
\[
C(P)=\frac{1}{2}\log_2\left(1+\frac{P}{N}\right).
\] Note that, if the utility function in ({\ref{eq:d_control}) is concave in distortion, the optimization problem in ({\ref{eq:d_control}) is in convex form\footnote{This is not the convex formulation in Section \ref{sec:RDframework}}. This follows from the fact that entropy is concave.

Next, for deriving further insights into the network control problem, we consider the case where utility $V_i(D_i)$ in ({\ref{eq:d_control}) is a linear function of $H(D_i)$, i.e., \[V_i(D_i) = -\delta_iH(D_i)\] for some constant $\delta_i > 0$. With change of variables $x_i = s_iH(D_i)$, from (\ref{eq:d_control}), we obtain an equivalent linear program (LP) (with sign of optimal value reversed) given by
\beq
\label{eq:d_control_lp}
&\min \limits_{x_1,x_2} & \frac{\delta_1}{s_1}x_1 + \frac{\delta_2}{s_2}x_2
\eeq
subject to
\beq
&& x_i \ge s_i H(p_i) - C(P_i), \forall i, \n \\
&& x_1+x_2 \ge s_1H(p_1) + s_2H(p_2)- C(P_1+P_2), \n \\
&&x_i \ge 0, x_i \le s_i, \forall i. \n 
\eeq

From properties of LP, it follows that at least one optimal solution exists that is a corner point of the feasible set, which is the convex polytope characterized by the constraints of the problem in (\ref{eq:d_control_lp}). More intuitively, we can obtain the optimal corner points for different cases based on where the source entropy vector $\bH=(s_1 H(p_1), s_2 H(p_2))$ lies with respect to the MAC capacity region $\cC$:
\begin{enumerate}
\item Case-A ($\bH \in \cC$): The optimal corner point is $D^*_1 = 0$, $D^*_2=0$, i.e., perform lossless source-coding.
\item Case-B ($\bH \notin \cC$): It follows from the MAC capacity region (and utility function) that there are only two corner points of interest. These are the corner points on the sum-capacity boundary. The exact corner points and the condition for choosing between these corner points are as follows: If ${\delta_1}/{s_1} \ge{\delta_2}/{s_2}$, then
\beq
s_1H(D^*_1) &=& \left[s_1H(p_1) - C(P_1)\right]^+, \n \\
s_2H(D^*_2) &=& \left[s_1H(p_1) - (C(P_1+P_2)-C(P_1))\right]^+, \n
\eeq
otherwise,
\beq
s_1H(D^*_1) &=& \left[s_1H(p_1) - (C(P_1+P_2)-C(P_2))\right]^+, \n \\
s_2H(D^*_2) &=& \left[s_1H(p_1) - C(P_2)\right]^+. \n
\eeq
\end{enumerate}
Here, $[x]^+$ denotes the positive part of $x$ given by $\max\{0,x\}$. 

Thus, we have explicitly solved the network control problem for this illustrative example. We depict this solution in Figure \ref{fig:r_opt_binary}. This figure captures the intuitive distortion-control policy: compute weights and choose the corner point for operation corresponding to the largest weight.

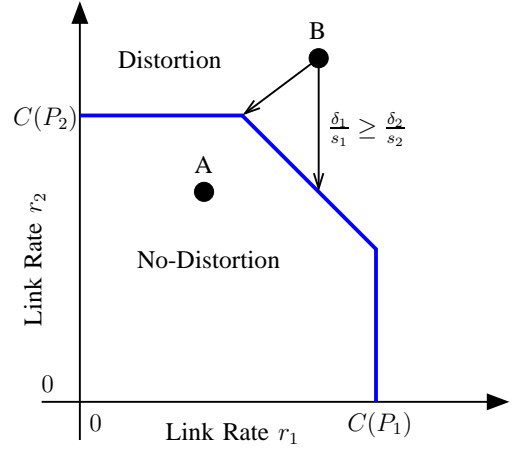
\begin{figure}[!h]
\centering
\scalebox{0.8}{\input{./r_opt_2ary.pstex_t}}
\caption{Optimal max-weight scheduling and distortion control for MAC with binary sources; Point-A corresponds to Case-A (no-distortion), and  Point-B corresponds to Case-B (distortion)}
\label{fig:r_opt_binary}
\end{figure}

\section{Conclusion}
\label{sec:concl}

We incorporate lossy compression into network control using rate-distortion theory. We do so as a user's happiness is heavily dependent on the distortion he or she observes in the lossy compression process. For sources such as video, such an optimization is especially relevant. We provide a provably optimal layered architecture for performing network control with minimal deviation from existing architecture. 


\end{document}

%% file: rd_control.pstex_t
\begin{picture}(0,0)%
\includegraphics{rd_control.pstex}%
\end{picture}%
\setlength{\unitlength}{3947sp}%
\begingroup\makeatletter\ifx\SetFigFont\undefined%
\gdef\SetFigFont#1#2#3#4#5{%
  \reset@font\fontsize{#1}{#2pt}%
  \fontfamily{#3}\fontseries{#4}\fontshape{#5}%
  \selectfont}%
\fi\endgroup%
\begin{picture}(4012,3766)(-239,-5980)
\put(1726,-3586){\makebox(0,0)[lb]{\smash{{\SetFigFont{12}{14.4}{\familydefault}{\mddefault}{\updefault}{\color[rgb]{0,0,0}No-Distortion}%
}}}}
\put(  1,-5311){\makebox(0,0)[lb]{\smash{{\SetFigFont{12}{14.4}{\familydefault}{\mddefault}{\updefault}{\color[rgb]{0,0,0}$0$}%
}}}}
\put(  1,-3886){\makebox(0,0)[lb]{\smash{{\SetFigFont{12}{14.4}{\familydefault}{\mddefault}{\updefault}{\color[rgb]{0,0,0}$1$}%
}}}}
\put(451,-5611){\makebox(0,0)[lb]{\smash{{\SetFigFont{12}{14.4}{\familydefault}{\mddefault}{\updefault}{\color[rgb]{0,0,0}$0$}%
}}}}
\put(1576,-5611){\makebox(0,0)[lb]{\smash{{\SetFigFont{12}{14.4}{\familydefault}{\mddefault}{\updefault}{\color[rgb]{0,0,0}$\frac{1}{K}$}%
}}}}
\put(2401,-5911){\makebox(0,0)[lb]{\smash{{\SetFigFont{12}{14.4}{\familydefault}{\mddefault}{\updefault}{\color[rgb]{0,0,0}Rate $c$}%
}}}}
\put(-224,-2986){\makebox(0,0)[lb]{\smash{{\SetFigFont{12}{14.4}{\familydefault}{\mddefault}{\updefault}{\color[rgb]{0,0,0}$\frac{H(D)}{H(p)}$}%
}}}}
\put(526,-3586){\makebox(0,0)[lb]{\smash{{\SetFigFont{12}{14.4}{\familydefault}{\mddefault}{\updefault}{\color[rgb]{0,0,0}Distortion}%
}}}}
\end{picture}%

%% file: r_opt_2ary.pstex_t
\begin{picture}(0,0)%
\includegraphics{r_opt_2ary.pstex}%
\end{picture}%
\setlength{\unitlength}{3947sp}%
\begingroup\makeatletter\ifx\SetFigFont\undefined%
\gdef\SetFigFont#1#2#3#4#5{%
  \reset@font\fontsize{#1}{#2pt}%
  \fontfamily{#3}\fontseries{#4}\fontshape{#5}%
  \selectfont}%
\fi\endgroup%
\begin{picture}(3937,3550)(-164,-5764)
\put( 76,-5311){\makebox(0,0)[lb]{\smash{{\SetFigFont{12}{14.4}{\familydefault}{\mddefault}{\updefault}{\color[rgb]{0,0,0}$0$}%
}}}}
\put(1276,-3586){\makebox(0,0)[lb]{\smash{{\SetFigFont{12}{14.4}{\familydefault}{\mddefault}{\updefault}{\color[rgb]{0,0,0}A}%
}}}}
\put(2176,-2536){\makebox(0,0)[lb]{\smash{{\SetFigFont{12}{14.4}{\familydefault}{\mddefault}{\updefault}{\color[rgb]{0,0,0}B}%
}}}}
\put(451,-5611){\makebox(0,0)[lb]{\smash{{\SetFigFont{12}{14.4}{\familydefault}{\mddefault}{\updefault}{\color[rgb]{0,0,0}$0$}%
}}}}
\put(2476,-5611){\makebox(0,0)[lb]{\smash{{\SetFigFont{12}{14.4}{\familydefault}{\mddefault}{\updefault}{\color[rgb]{0,0,0}$C(P_1)$}%
}}}}
\put(-149,-3211){\makebox(0,0)[lb]{\smash{{\SetFigFont{12}{14.4}{\familydefault}{\mddefault}{\updefault}{\color[rgb]{0,0,0}$C(P_2)$}%
}}}}
\put(1051,-5686){\makebox(0,0)[lb]{\smash{{\SetFigFont{12}{14.4}{\familydefault}{\mddefault}{\updefault}{\color[rgb]{0,0,0}Link Rate $r_1$}%
}}}}
\put( 76,-4786){\rotatebox{90.0}{\makebox(0,0)[lb]{\smash{{\SetFigFont{12}{14.4}{\familydefault}{\mddefault}{\updefault}{\color[rgb]{0,0,0}Link Rate $r_2$}%
}}}}}
\put(2326,-3286){\makebox(0,0)[lb]{\smash{{\SetFigFont{12}{14.4}{\familydefault}{\mddefault}{\updefault}{\color[rgb]{0,0,0}$\frac{\delta_1}{s_1}\ge \frac{\delta_2}{s_2}$}%
}}}}
\put(676,-2761){\makebox(0,0)[lb]{\smash{{\SetFigFont{12}{14.4}{\familydefault}{\mddefault}{\updefault}{\color[rgb]{0,0,0}Distortion}%
}}}}
\put(826,-4336){\makebox(0,0)[lb]{\smash{{\SetFigFont{12}{14.4}{\familydefault}{\mddefault}{\updefault}{\color[rgb]{0,0,0}No-Distortion}%
}}}}
\end{picture}%